# Sensing the Pulse of the Pandemic: Geovisualizing the Demographic Disparities of Public Sentiment toward COVID-19 through Social Media

Binbin Lin[a], Lei Zou[a]*, Bo Zhao[b], Xiao Huang[c], Heng Cai[a], Mingzheng Yang[a], and Bing Zhou[a]

[a] *Department of Geography, Texas A&M University, College Station, USA*

[b] *Department of Geography, University of Washington, Seattle, USA*

[c] *Department of Environmental Sciences, Emory University, Atlanta, USA*

*\* lzou@tamu.edu*

# Sensing the Pulse of the Pandemic: Geovisualizing the Demographic Disparities of Public Sentiment toward COVID-19 through Social Media

Social media offers a unique lens to observe large-scale, spatial-temporal patterns of users' reactions toward critical events. However, social media use varies across demographics, with younger users being more prevalent compared to older populations. This difference introduces biases in data representativeness, and analysis based on social media without proper adjustment will lead to overlooking the voices of digitally marginalized communities and inaccurate estimations. This study explores solutions to pinpoint and alleviate the demographic biases in social media analysis through a case study estimating the public sentiment about COVID-19 using Twitter data. We analyzed the pandemic-related Twitter data in the U.S. during 2020-2021 to (1) elucidate the uneven social media usage among demographic groups and the disparities of their sentiments toward COVID-19, (2) construct an adjusted public sentiment measurement based on social media, the Sentiment Adjusted by Demographics (SAD) index, to evaluate the spatiotemporal varying public sentiment toward COVID-19. The results show higher proportions of female and adolescent Twitter users expressing negative emotions to COVID-19. The SAD index unveils that the public sentiment toward COVID-19 was most negative in January and February 2020 and most positive in April 2020. Vermont and Wyoming were the most positive and negative states toward COVID-19.

## 1. Introduction

The global COVID-19 pandemic has caused diverse, severe mental stresses to different social groups, leading to adverse psychological and physical health impacts. According to a scientific brief released by the World Health Organization (WHO), COVID-19 triggered a 25% increase in the prevalence of anxiety and depression worldwide in 2020 (Brunier, 2022). This phenomenon was also observed in the U.S., where symptoms of anxiety and depressive disorders increased significantly from April to June 2020, compared with the same period in 2019 (Czeisler et al., 2020). Therefore, understanding the spatial and temporal disparities of human emotional stresses under the pandemic's impacts, e.g., travel restrictions and health consequences, is of great importance that necessitates investigations.

Social media are online communication platforms that facilitate people across the globe to access information on "what's happening," share perspectives and emotions, and voice attitudes and reactions to real-world events in the virtual space (Lin et al., 2022; Zhou et al., 2022; Zou et al., 2018). Data collected from social media platforms offer a unique lens to observe users' digital traces that reflect their opinions and subjective feelings at fine spatial-temporal scales in a low-cost and efficient manner. These large amounts of human-generated data are difficult to obtain from traditional survey methods (Alamoodi et al., 2021). Therefore, analyzing social media data makes it possible to uncover the spatial-temporal disparities of public emotions toward COVID-19 and offer valuable insights into the effects of the pandemic on mental stresses.

However, the data bias issue is a persistent challenge in research based on data collected from social media. Existing evidence uncovers that the usages of Twitter, one of the most popular social media platforms, are biased toward younger, well-educated, and wealthier populations living in urban communities (Blank 2017; Jiang, Li, and Ye 2018). Consequently, analyzing human sentiment using social media data without considering the underlying demographic biases may overlook the stress of certain social groups and lead to unfair estimations (Hung et al., 2020). Since most social media platforms do not provide demographic information about their users, it is difficult to investigate and remediate the uneven user distribution in social media-based research. Nevertheless, the advance in artificial intelligence (AI) models has led to new solutions to decipher social media users' demographics (Wang et al., 2019). Leveraging those advanced models is promising in revealing the social media activities of different social groups and yielding demographically adjusted estimations of public sentiment.

This study proposes to incorporate AI-empowered demographic inference models to uncover the demographic disparities in social media usage and mitigate the demographic bias issue in social media analysis. The proposed method is demonstrated through a case study estimating public sentiments relevant to the COVID-19 pandemic. We collected and analyzed Twitter data in the U.S. from January 2020 to December 2021 to answer two research questions. First, what are the geographical, temporal, and demographic disparities of public sentiment toward COVID-19 reflected on social media? Second, how can we alleviate the demographic bias within social media to fairly evaluate spatial and temporal patterns of public sentiment toward specific topics or events like COVID-19? To address these research questions, three objectives were proposed and achieved: (1) to elucidate the uneven social media usage among various demographic groups and the disparities of their emotions toward COVID-19 in the U.S., (2) to construct an adjusted public sentiment measurement based on social media data, the Sentiment Adjusted by Demographics (SAD) index, through the post-stratification method, and (3) to evaluate the spatially and temporally evolved public sentiment toward COVID-19 at the state level using the SAD index. The overarching hypothesis is that female and adolescent users expressed more negative sentiment toward COVID-19 on Twitter in the U.S. Evidence supporting this hypothesis has been discovered in small-size sampled, individual-based psychological studies (G. Knowles et al., 2022; Liu et al., 2020; Terry et al., 2020), but whether this hypothesis still holds when considering large-scale social groups across space and during evolving pandemic phases is unclear and requires further investigations. Through revealing the demographic biases within social media usage and data, the voices of digitally marginalized communities will be identified and incorporated into estimating the public sentiment during the pandemic. The results will support governments in developing evidence-based strategies to mitigate depressive emotions for different populations. The proposed framework can be used to tackle the demographic bias challenge in social media analytics for similar studies.

## 2. Related work

### *2.1 Demographic bias in Twitter data*

Twitter data are widely acknowledged as one of the most popular social media datasets for investigating human perceptions and dynamics at unprecedented spatial and temporal scales. Concurrently, concerns regarding the dependability of social media data in capturing human behaviors and attitudes have emerged, mainly due to the heterogeneous user composition found across various social media platforms. Previous efforts have demonstrated that Twitter users are demographically biased, and human behaviors observed from Twitter could not directly represent the activities of the general population (Barbera, 2016; Cesare et al., 2019; Jiang et al., 2018; Sloan et al., 2015).

Researchers have sought to reveal the innate bias of Twitter users using two approaches. The first approach leverages information in Twitter user profiles to detect each user's demographic information, such as age, gender, and ethnicity, and compares users' activities by those demographic attributes. A common method to infer the demographics of social media users is by analyzing their names. For instance, Mandel et al. (2012) examined the public sentiment in the U.S. during 2011 Hurricane Irene and its gender differences. They collected hurricane-related Twitter data along with the corresponding user profiles and inferred users' genders by matching their names with a list of common male and female names from the Census Bureau. The results show that female users were more likely to express concerns on social media during emergencies than male users. Longley and Adnan (2016) inferred Twitter users' age and ethnicity based on their given and family names, which together with users' geo-temporal activities, were used to cluster users in London into seven groups: London residents, students, visitors, spectators of major events, commuting professionals, the daily grind, and workers and tourists. Users' tweet content and social networks can also be utilized to estimate their sociodemographic characteristics. In Barbera (2016), a machine learning model was calibrated to predict users' ages, genders, and races with the training dataset from 250,000 Twitter users with voting registration records. The study discovers that female users prefer to use motion words in tweets, and people over 40 are more likely to post messages about children or grandchildren.

Another approach employs spatial-temporal models to investigate the relationships between the density of Twitter users or messages and the demographic structure of the population to illustrate the Twitter data bias. Jiang et al. (2018) applied the Geographically Weighted Regression (GWR) model to infer the relationships between three demographic and two socioeconomic factors and the number of geotagged Twitter users in the U.S. at the county level. The results show that the 18–29 age group has the most active geotagged Twitter users, and the percentage of well-educated populations (people with a Bachelor or higher degree) tends to have a stronger positive relationship with the number of Twitter users. A study in California explored the correlations between tweet density and the socioeconomic characteristics of local residents at the county level (L. Li et al., 2013). They found that well-educated, higher-income people working in business, science, arts, and management were more likely to generate dense geo-referenced Twitter data.

*2.2 Sentiment analysis of COVID-19-related Twitter data*

Sentiment analysis is a natural language processing (NLP) technique that quantifies affective states and subjective feelings (Hasan et al., 2019). Based on textual data such as tweet content, sentiment analysis can detect the emotions or attitudes embedded in texts at a simple dual polarity-based scale or a multidimensional emotion spectrum (Giatsoglou et al., 2017) through lexicon-based or machine learning-based approaches (Khan et al., 2020). While human emotions are undeniably intricate and cannot be fully captured by a single emotion or numerical value, sentiment analysis still holds significant value. It provides valuable insights into the overall emotional tone of a text, allowing us to gain meaningful understanding from vast amounts of textual data. Since the COVID-19 outbreak worldwide, numerous studies have leveraged different sentiment analysis methods and tools to investigate the attitudes and emotions of tweets relevant to COVID-19 at various geographical regions and scales.

Boon-Itt and Skunkan (2020) collected and analyzed COVID-19-related English tweets from December 13th, 2019 to March 9th, 2020. They applied the National Research Council (NRC) sentiment lexicon to categorize tweets into ten basic emotions, i.e., anger, anticipation, disgust, fear, joy, negative, positive, sadness, surprise, and trust. The results show that over half of the tweets' emotions were classified as fear, trust, and anticipation, with fear being the most common sentiment. Hung et al. (2020) evaluated the sentiment of COVID-19-related English tweets between March 20th to April 19th, 2020, in the U.S. by a Python package named Valence Aware Dictionary and sEntiment Reasoner (VADER), which returns sentiment scores from -1 to 1 at a polarity-based scale. They found that 48.2% of tweets expressed a positive sentiment, while 31.1% showed negative emotions. Commonly used neural network models for context understanding, including the long short-term memory (LSTM) and bidirectional LSTM (Bi-LSTM) methods, were also applied to estimate sentiment scores of tweets concerning COVID-19 vaccination from December 21st to July 21st, 2020, at the global scale (Alam et al., 2021). The results reveal that the majority of the tweets held a neutral sentiment, and the negative tweets were the fewest. Bokaee, Nezhad, and Deihimi (2022) identified sentiments in persian tweets discussing COVID-19 vaccination between April 1st, 2021, and September 30th, 2021, to investigate Iranian Twitter users' views toward homegrown and foreign vaccines. A customized model concatenating

Convolutional Neural Networks (CNN) with LSTM was built for sentiment classification. They concluded that the percentage of positive sentiments for foreign vaccines was slightly higher than for domestic vaccines.

In sum, extracting sentimental information from Twitter data has become a popular approach to understanding human perceptions toward COVID-19 in near real-time. However, few studies explore the disparities of public sentiment about COVID-19 among different social groups considering the demographic bias of Twitter data and propose practical solutions to alleviate such user unevenness. This research is poised to address this challenge.

**3. Data and Method**

This study utilized Twitter data and Census data from the U.S. to uncover the demographic bias in social media data and examine public sentiment toward COVID-19. The process, depicted in Figure 1, consists of three main steps. The first step is Twitter data mining for sentiment evaluation, including Twitter data collection, cleaning, and sentiment estimation. Second, Twitter users' demographics were identified to group users by age and gender and determine any disparities in their sentiments regarding COVID-19. Finally, we applied the post-stratification algorithm to compute the Sentiment Adjusted by Demographics (SAD) index to yield an adjusted evaluation of the sentiment. The details of each step are illustrated from Section 3.1 to 3.3.

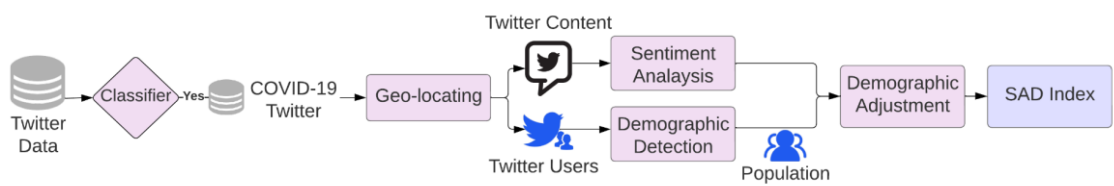

Figure 1. Framework of Twitter data mining for demographic-adjusted public sentiment evaluation.

*3.1 Twitter data mining for sentiment evaluation*

We collected Twitter data in the U.S. from January 2020 to December 2021 through the Twitter Application Programming Interface (API) for Academic Research (https://developer.twitter.com/en/products/twitter-api/academic-research). The Twitter Academic API enables researchers to access complete Twitter datasets up to 10 million tweets per month with a streaming rate limit of 50 requests every 15 minutes for each approved account. Two sets of keywords were selected to identify COVID-19-related English tweets in the U.S. based on an overview of previous literature (Alqurashi, Alhindi, and Alanazi 2020; Chen, Chen, and Pang 2022; Lin et al. 2022). As shown in Table 1, 14 keywords were used for Twitter data collection in 2020 and the list was expanded to include 15 vaccination-related keywords for data collection in 2021. Any English tweet containing at least one of these keywords was collected in the initial database. Tweets posted by bots and organizational users were considered irrelevant to public discussion about the coronavirus and removed from the database by examining the 'source' and 'verified' attributes in Twitter data (Lin et al., 2022). Specifically, a list of source (e.g., sent from iPad) created by Li et al. (2021) were employed to identify human-generated tweets. 'Verified' Twitter users were removed since they are usually accounts of non-general-public, e.g., government and organizations. A total of 4,822,802 geotagged tweets located in the U.S. were collected after Twitter data cleaning.

Table 1. Keywords used for Twitter data collection in 2020 and 2021.

| Year | Keyword Lists | References |
|---|---|---|
| 2020 | covid, virus, 2019-ncov, sars-cov-2, coronavirus, ncov, n95, social distancing, lockdown, quarantine, pandemic, epidemic, pneumonia, confirmed cases | Alqurashi et al., 2020; Lin et al., 2022 |
| 2021 | covid, virus, 2019-ncov, sars-cov-2, coronavirus, ncov, n95, social distancing, lockdown, quarantine, pandemic, epidemic, pneumonia, confirmed cases, vax, vaccin, covidvic, impfstoff, | Chen et al., 2022 |

> vacin, vacuna, impfung, sputnikv, astrazeneca, sinovac, pfizer,
> moderna, janssen, johnson, biontech

Finally, we evaluated the sentiment of each tweet based on the text content using the VADER package proposed by Hutto and Gilbert (2014). VADER is a NLP-based sentiment analysis tool. It provides an index ranging from –1 to 1, with –1 representing the most negative sentiment and 1 representing the most positive sentiment. The VADER sentiment analysis algorithm was recently updated in April 2022 to ensure its effectiveness in capturing evolving expressions of sentiment in online communication.

*3.2 Detection of Twitter Users' Demographics*

This study employed the M3 (multimodal, multilingual, and multi-attribute) model proposed by Wang et al. (2019) to infer two demographic characteristics of Twitter users, i.e., age (≤18, 19-29, 30-39, or ≥40 years of age) and gender (male or female), based on users' screen names, usernames, profile images, and biographies. The M3 model training and prediction did not incorporate a more comprehensive age and gender classification. As a result, this investigation will focus solely on the four age groups and two genders identifiable by this module. The open-source package of the M3 model was obtained from GitHub (https://github.com/euagendas/m3inference). The M3 model combines the strengths of a DenseNet model, a dense convolutional network model, and a 2-stack bidirectional character-level Long-short Term Memory (LSTM) model to extract information from profile images and text data (username and biography). This model was trained on massive, various types of datasets, including Twitter data in which gender and age were identified from expression in the biographies, e.g., "Mother of a wonderful kid" or "Happy 24[th] birthday", a curated dataset of Twitter accounts belonging to organizations, face images from IMDB and Wikipedia, and a crowdsourced dataset in 32 languages. The accuracies of the M3 model in gender and age recognition in the English test dataset are 0.81 and 0.42, respectively, higher than other existing models such as those in Jaech and Ostendorf (2015), Knowles, Carroll, and Dredze (2016), and Wang and Jurgens (2018).

### *3.3 Sentiment Adjusted by Demographics (SAD) Index*

The widely adopted method to measure aggregated sentiment is taking average sentiment scores of all tweets (Chatterjee & Perrizo, 2016; Guzman & Maalej, 2014; Scrivens et al., 2015). However, this method may lead to inaccurate sentiment evaluation. For example, an average sentiment score of 0 could be the mean value of two tweets with their sentiment scores as -1 and 1. As a result, the average score indicates a neutral sentiment but overlooks the disparate emotions of individual users and messages. To address the issue, this study defines the Sentiment index as the weighted proportion of Twitter users within a specific community and time frame who exhibit overall negative emotions. Initially, each Twitter user's sentiment is calculated as the average sentiment of all COVID-19-related tweets posted by this user within a given timeframe. Afterwards, we divided the continuous sentiment scores of individual users into three categories: negative (-1~-0.05), neutral (-0.05~0.05), and positive (0.05~1) (Bonta et al., 2019; Mardjo & Choksuchat, 2022).

Due to the demographic bias in Twitter data, sentiment evaluation directly derived from Twitter data, e.g., the percentage of users expressing negative emotions, is inaccurate in representing the public sentiment of the whole population. Hence, we employed the post-stratification method, a weighting adjustment method commonly used in survey data processing, to alleviate the deviation caused by demographic bias (David A. Dargatz & G. W. Hill, 1996; Kalton & Flores-Cervantes, 2003; Vaske et al., 2011). The adjustment weighting was calculated based on the demographic discrepancy between Twitter users and the general population using Equation 1:

$$w_{ij} = \frac{p_i(\text{population})}{p_{ij}(\text{Twitter users})} \quad (1)$$

where $i$ denotes the $i$th of the eight social groups classified by age and gender, $j$ denotes the time period, $p_i(\text{population})$ represents the proportion of the $i$th social group in the whole population, $p_{ij}(\text{Twitter users})$ denotes the proportion of the $i$th social group in Twitter users during time period $j$, and $w_{ij}$ means the weight for $i$th social group in $j$ time period. The population demographic data in 2020 was collected from the U.S. Census Bureau and used to adjust the bias within social media data (Bureau, 2021).

Afterwards, we counted the numbers of Twitter users in different sentiment levels (negative neutral, and positive) in the time span $j$, and utilized the weights derived from Equation 1 to adjust the numbers of Twitter users in each sentiment category.

$$\text{Adjusted percentage}_{lj} = \frac{\sum_{i=1}^{8} N_{lij} w_{ij}}{\sum_{i=1}^{8}((\sum_{l=1}^{3} N_{lij}) w_{ij})} \quad (2)$$

In Equation 2, $i$ represents one of the eight social groups, while $j$ denotes the analysis timeframe. The variable $l$ specifies the sentiment levels, which are categorized as negative (1), neutral (2), and positive (3). $N_{lij}$ corresponds to the number of Twitter users within social group $i$ who exhibit an overall sentiment level $l$ during timeframe $j$. The calculated outcome, denoted as $\text{Adjusted percentage}_{lj}$, represent the overall percentage of Twitter users displaying sentiment level $l$ during timeframe $j$ after accounting for demographic biases. Specifically, the Sentiment Adjusted by Demographics (SAD) Index for a given timeframe $j$ is defined as the percentage of Twitter users expressing an overall negative emotion toward COVID-19 during the period $j$ after adjusting for demographic characteristics. The python code example of the post-stratification method is available in a GitHub repository (https://github.com/binbinlinGISer/Sentiment-adjusted-by-demographics-SAD-Index).

## 4. Results

*4.1 Demographic bias of Twitter users*

Figure 2 illustrates the demographic disparities between Twitter users who posted messages about COVID-19 and the actual population in the U.S. regarding gender and age. The proportion of Twitter users under 18 is 14.03%, which is smaller than the corresponding population segment in the U.S. (23.36%). The age group between 19 and 29 accounts for 35.35% of Twitter users, more than double the proportion found in the U.S. population (14.90%). The rate of Twitter users from 30 to 39 is 18.76%, surpassing the percentage of individuals in this age range within the U.S. population (13.56%). Notably, there exists a significant discrepancy between the proportions of females over 40 on Twitter (9.43%) and in the U.S. population (25.31%). These differences align with previous studies that have also reported a concentration of younger cohorts and a higher proportion of male users on Twitter (Cesare et al., 2019; Filho et al., 2015; Jiang et al., 2018; Longley et al., 2015). This observation quantitively demonstrates the presence of demographic bias in Twitter data. It highlights the need to adjust this bias when assessing sentiment from Twitter data to avoid over- or under-representing certain social groups.

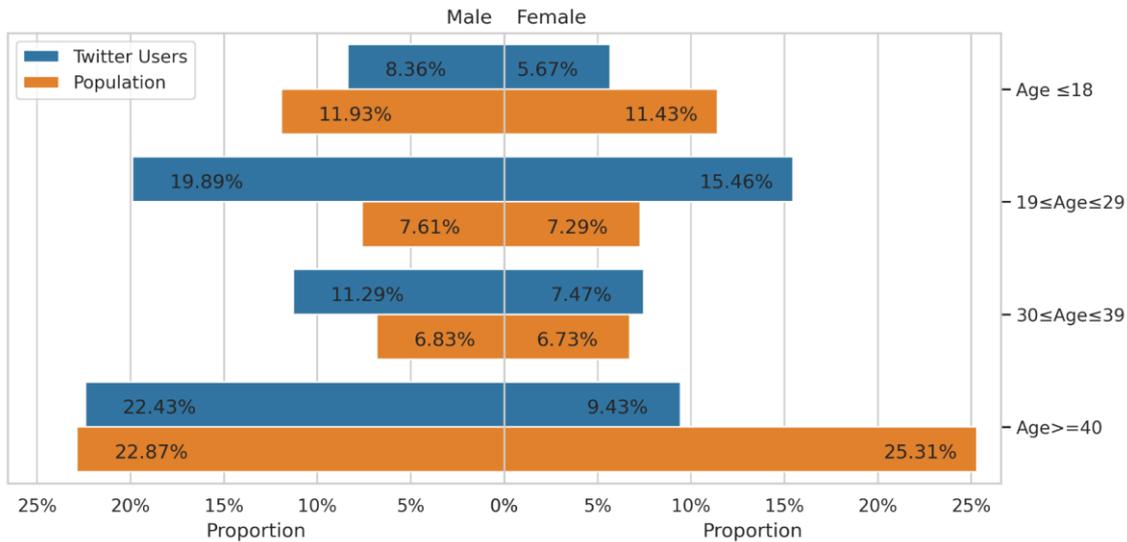

Figure 2. Comparison of demographic structures between Twitter users discussing COVID-19 and the actual population in the U.S.

*4.2 Demographic disparities of public sentiment toward COVID-19 in the U.S.*

Table 2 lists four instances of COVID-19-related tweets showcasing extreme sentiment across various demographic groups. Two example tweets with a sentiment score lower than -0.8 were authored by a female under 18 years old and a male between 30 and 39 years old, as indicated by demographic detection. These tweets focused on the topics of COVID-19 testing and symptoms. Conversely, the other two tweet examples with a sentiment score higher than 0.9 were posted by a male between 19 and 29 years old and a female over 40 years old. These tweets celebrated reunions and acknowledged the value of COVID-19 policies.

Table 2. Instances of COVID-19 related tweets displaying extreme sentiment in different demographic groups.

| Twitter content | M3 Predicted Age | M3 Predicted Gender | Sentiment Score | Location (State) |
| --- | --- | --- | --- | --- |
| and idk if it's covid cause i'm still waiting for the results but just to be safe im staying home 😭😭😭 | Age≤18 | Female | -0.87 | California |
| Yesterday we had a blast! 13months way to long to see our good pal JD perform and haven't seen these guys since the beginning of this pandemic was worth the wait! Love these faces! 💞 | 19≤Age≤29 | Male | 0.93 | California |
| health department called and interviewed me about my covid experience. Main symptoms I've had are fever, fatigue, shivering, body aches, headaches, congestion, joint pain, and horrific insomnia. I miss sleeping 😭 | 30≤Age≤39 | Male | -0.94 | Rhode Island |
| Lockdown does save precious lives 😷🖤 | Age≥40 | Female | 0.90 | Texas |

Figure 3 shows the percentages of Twitter users in three COVID-19-related sentiment levels by age and gender in the United States within the initial two years of the pandemic. Three orange, grey, and blue octagons represent percentages of negative, neutral, and positive sentiments in each social group, with a total sum of 100% for each group. The rates of users expressing positive sentiments range from 44.90% to 60.62%, while the proportions of users expressing an overall negative emotion are comparatively lower, ranging from 23.72% to 33.25%.

The three octagons reveal the existence of disparities in COVID-19 sentiment across various social groups during the period of 2020 to 2021. Among the eight social groups analyzed, female Twitter users aged between 30 and 39 exhibited the highest proportion of positive users (60.62%) and the smallest percentage of negative users (23.72%). This group emerged as the most positive user group toward COVID-19 during the first two years of the pandemic. The Female & Age≥40 group followed behind as the second most positive social media user group, with 58.08% expressing positive sentiments and 25.89% expressing negative sentiments on Twitter. Regarding male users, those over 40 displayed the highest proportion of positive users (55.77%) and the lowest percentage of negative users (25.99%), followed by male users aged between 30 and 39. These observations indicate that Twitter users over 30 showed a more positive sentiment toward COVID-19 than other age groups during the investigated period.

On the contrary, 34.82% of female Twitter users below 18 expressed an overall negative emotion toward COVID-19. This is the highest among all social groups, followed by the male users below 18 (33.25%). The proportion of positive Twitter users was the smallest for males below 18 (44.90%), followed by females below 18 (46.21%). These phenomena suggest that Twitter users under 18 were mentally more sensitive to the pandemic and posted more negative messages regarding the virus on social media.

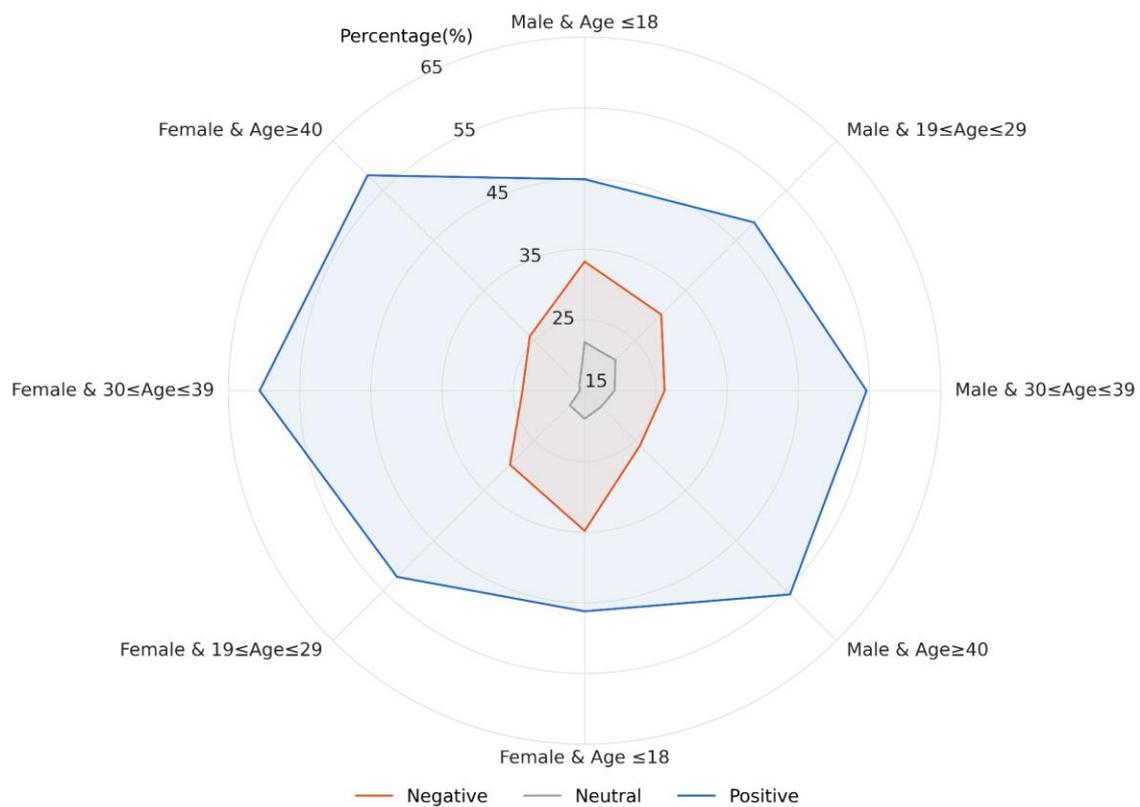

Figure 3. Percentages of negative, neutral, and positive Twitter users in eight social groups in the U.S.

Figure 4 depicts the monthly temporal trends in the proportions of Twitter users exhibiting depressive sentiments toward COVID-19 across various age and gender groups in the U.S. from 2020 to 2021. In general, the monthly percentages of negative Twitter users in different social groups displayed similar patterns. The values ranged from 23.54% to 45.19% with three peaks observed in February 2020, June 2020 to January 2021, and August to September 2021. Additionally, two troughs were detected in April 2020 and March to May 2021.

In terms of gender groups, the monthly percentages of negative Twitter users were generally higher among females compared males, with exceptions observed during the months of March to May and August in 2020. It reveals that, for the majority of the two years period, female users displayed slightly higher emotional vulnerability toward COVID-19 compared to male users. Regarding the sentiment disparities across age groups, the subfigure below shows that the Age≤18 group had the highest percentage of negative users throughout the two years.

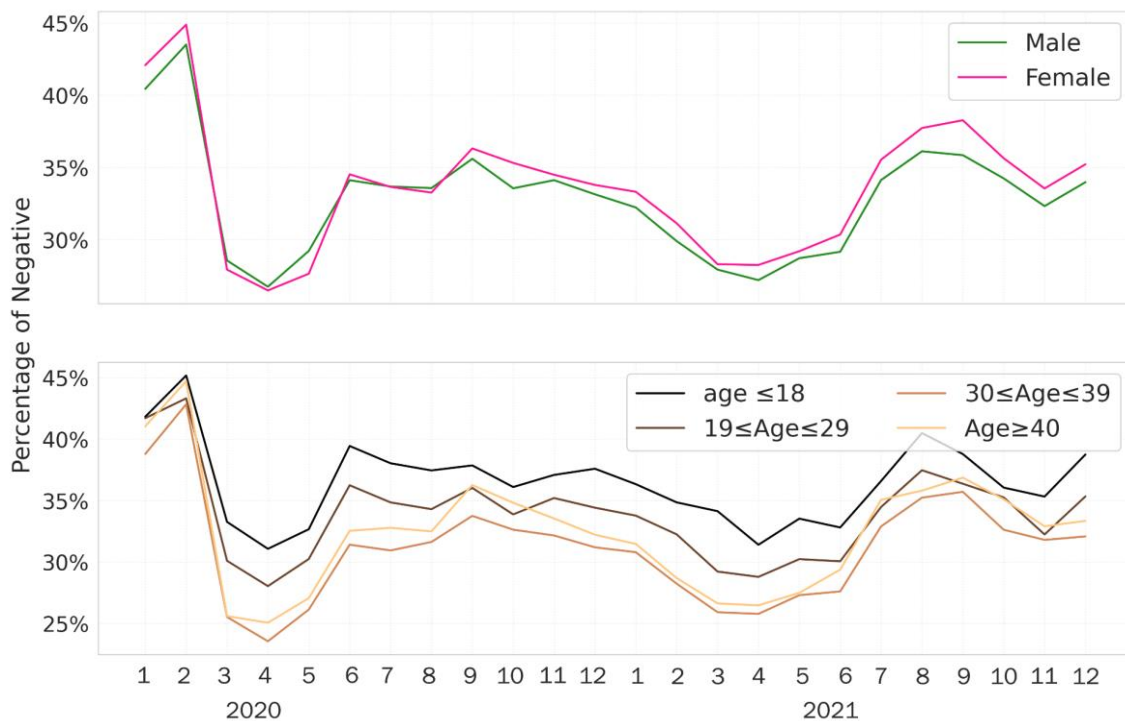

Figure 4. Temporal trends of percentages of negative Twitter users in different ages and gender groups in the U.S.

*4.3 Demographic and geographical disparities of public sentiment toward COVID-19*

The demographic composition and geographical distribution of users who tweeted about COVID-19 in the U.S. between 2020 and 2021 is uneven, as evident in Figure 5. Among the states, California (CA) had the highest number of Twitter users posting at least one geotagged tweet about COVID-19 (125,989), followed by Texas (TX, 86,142), New York (NY, 66,397), and Florida (FL, 51,838). On the other hand, Wyoming (WY) had the smallest number of Twitter users mentioned COVID-19 in their tweets (824), followed by North Dakota (ND, 1,256), Delaware (DE, 1,328), and South Dakota (SD, 1,375).

In most states, male users over 40 years old constituted the largest user group across all social groups, except in Vermont (VT), Mississippi (MS), Louisiana (LA), Georgia (GA), Texas, and California, where the largest group was male users between 19 and 29 years old. Across most states, the percentage of Twitter users in the Female & Age≤18 group posting about COVID-19 was the smallest, except in Hawaii (HI), Louisiana, and Texas, where the Female & 30≤Age≤39 group constituted the smallest group posting messages about COVID-19. It is worth noting that male users outnumbered female users in all age groups, particularly in the users over 40 years old.

In Figure 5, we also present the ratios of Twitter users who tweeted about COVID-19 within each social group, normalized by its corresponding population, revealing the state-level uneven representativeness of Twitter data without proper adjustments. The calculated ratio values varied from 0.04% to 1.29% (4 to 129 out of 10 thousand people posted at least one tweet about COVID-19). Remarkably, the highest ratio values were predominantly observed among males aged 19 to 29, particularly in states along the coast, followed by females in the same age segment. Conversely, the lowest values were found in females aged over 40, encompassing most states in the U.S. Overall, Twitter data demonstrated higher representativeness for males across all age groups compared to females.

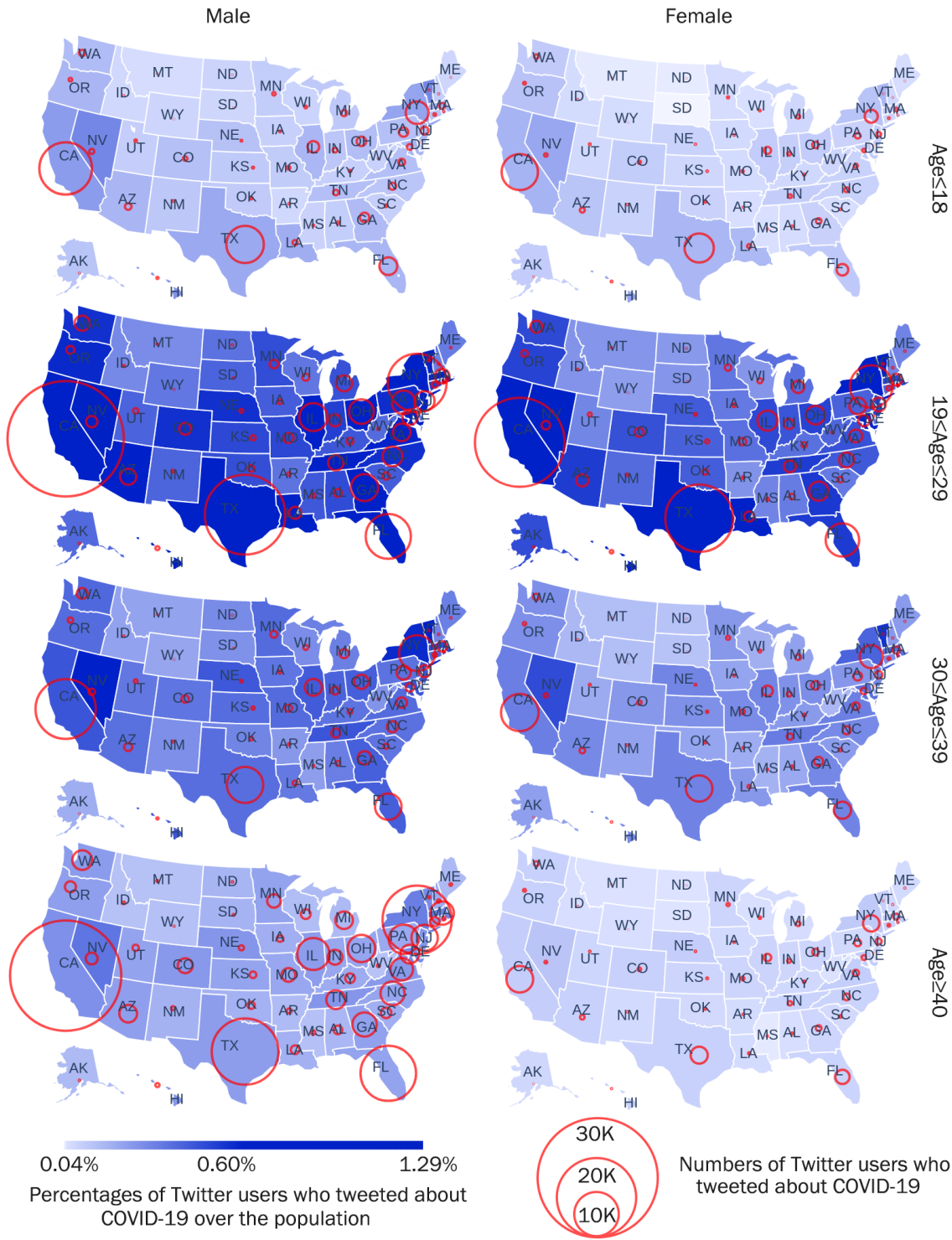

Figure 5. Numbers and ratios of Twitter users who tweeted about COVID-19 by gender and age at the state level in the U.S. during the period of 2020 to 2021.

Figure 6 illustrates the geographical and demographic disparities in public sentiment at the state level across the U.S. during 2020 to 2021. The percentages of Twitter users displaying an averaged negative sentiment toward COVID-19 ranged from 14.29% to 50%, with a mean value of 29.36%. In terms of geographical disparities, the highest average percentage of negative Twitter users was observed in Wyoming (WY) at 34.10%. This was followed by Alaska (AK), Montana (MT), and South Dakota (SD), with average proportions of 32.83%, 32.69%, and 31.74%, respectively. Wyoming's high proportion of negative Twitter users could be attributed to the state's small number of Twitter users and low representativeness, as explained earlier in Figure 5. Conversely, Vermont (VT) exhibited the smallest proportion of negative users on Twitter, with an average percentage of 21.25%. This was followed by Massachusetts (MA), Minnesota (MN), and Tennessee (TN) with average percentages of 27.09%, 27.44%, and 27.44% respectively.

The results also show that, in 37 states, the Female & Age $\leqslant$ 18 social group had the highest rate of Twitter users posting negative tweets about COVID-19. In 12 states, the Male & Age $\leqslant$ 18 group had the most proportion of Twitter users posting negative tweets. In Wyoming, for instance, female Twitter users under 18 exhibited the highest proportion of negative users, with 50% expressing negative sentiment. The Male & Age $\leqslant$ 18 group ranked as the second most negative social group in Wyoming, with 38.81% of Twitter users displaying negative sentiments. In Montana, the proportion of negative Twitter users in Male & Age $\leqslant$ 18 group was 41.57%, the highest among all social groups in this state.

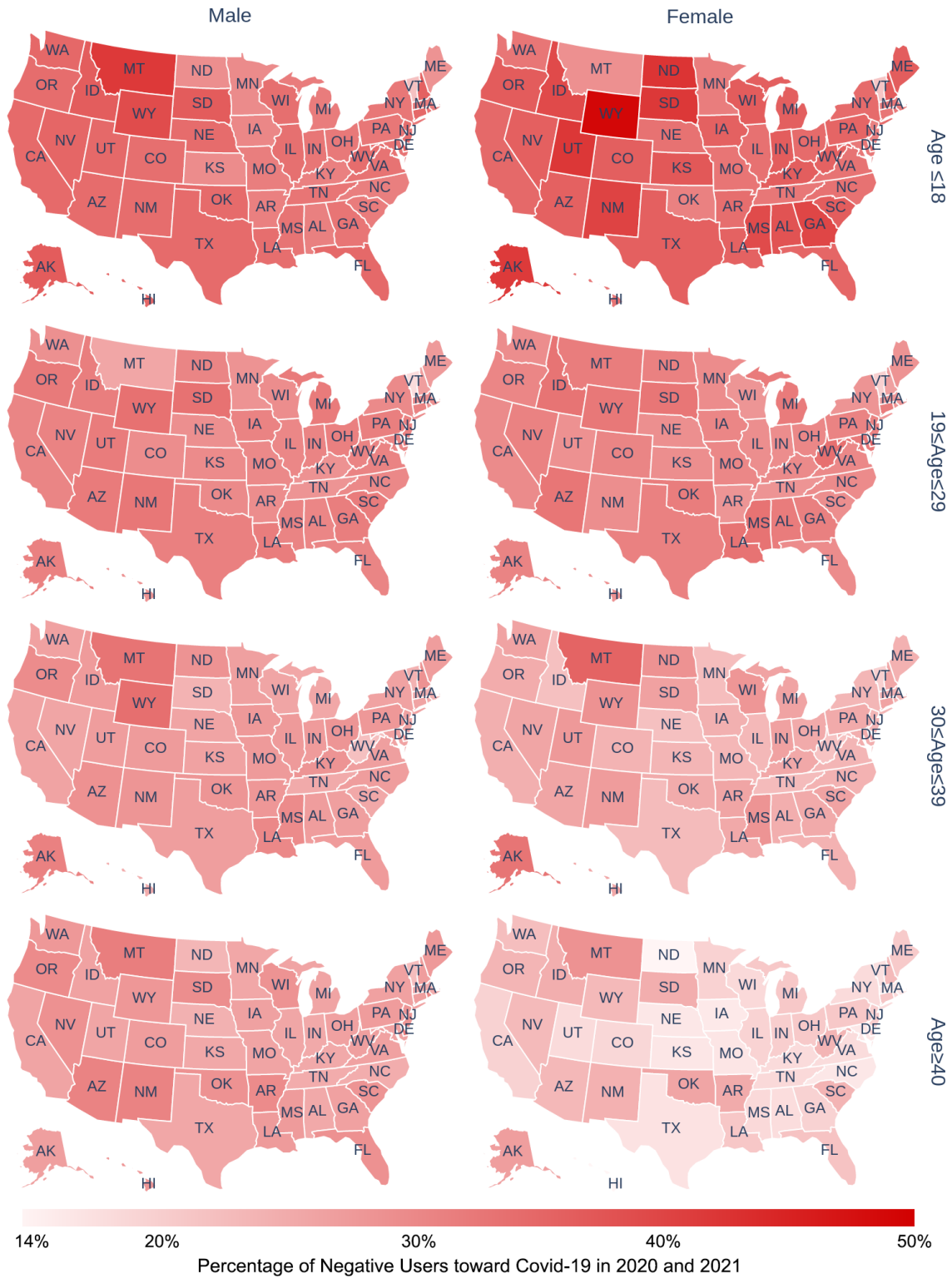

Figure 6. Percentages of negative Twitter users toward COVID-19 within various social groups at the state level across the U.S. during the period of 2020 to 2021.

Figure 7 depicts the monthly variations in the percentages of Twitter users expressing negative emotions toward COVID-19 in the six most populous states, including California,

Texas, Florida, New York, Pennsylvania, and Illinois. The data are presented as heatmaps, highlighting the monthly changes for each social group. In the heatmaps, grey cells indicate insufficient data (less than 50 users) to calculate the percentage of negative Twitter users for the corresponding social group in a specific month and state. The monthly percentages of negative Twitter users ranged from 20.28% to 55.56%, with an average value of 33.59%. Across the six states, Twitter users in the Female & Age ≤18 group consistently displayed higher levels of negative emotions in their COVID-19-related tweets compared to the other seven social groups throughout most months in 2020 and 2021.

Furthermore, the average proportions of negative Twitter users across the eight social groups were highest in January and February 2020 in the six states. During those months, the average percentages of negative users were 44.11%, 43.05%, 43.57%, 43.00%, 42.56%, and 43.29%, respectively. In addition, the Female & Age ≤18 group in California exhibited a high proportion (47.26%) of negative Twitter users in December 2021. Similarly, in Florida, the percentages of negative Twitter users were high in September 2021 for the Female & Age ≥40 group (47.00%) and the Female & 30≤Age≤39 group (44.83%). In New York, the Female & Age ≤18 group showed elevated percentages of negative Twitter users in August and September 2021 (46.06% and 47.45%). Lastly, in Illinois, the Female & Age ≤18 group displayed high proportions of negative Twitter users in December 2020 (45.33%) and December 2021 (44.64%).

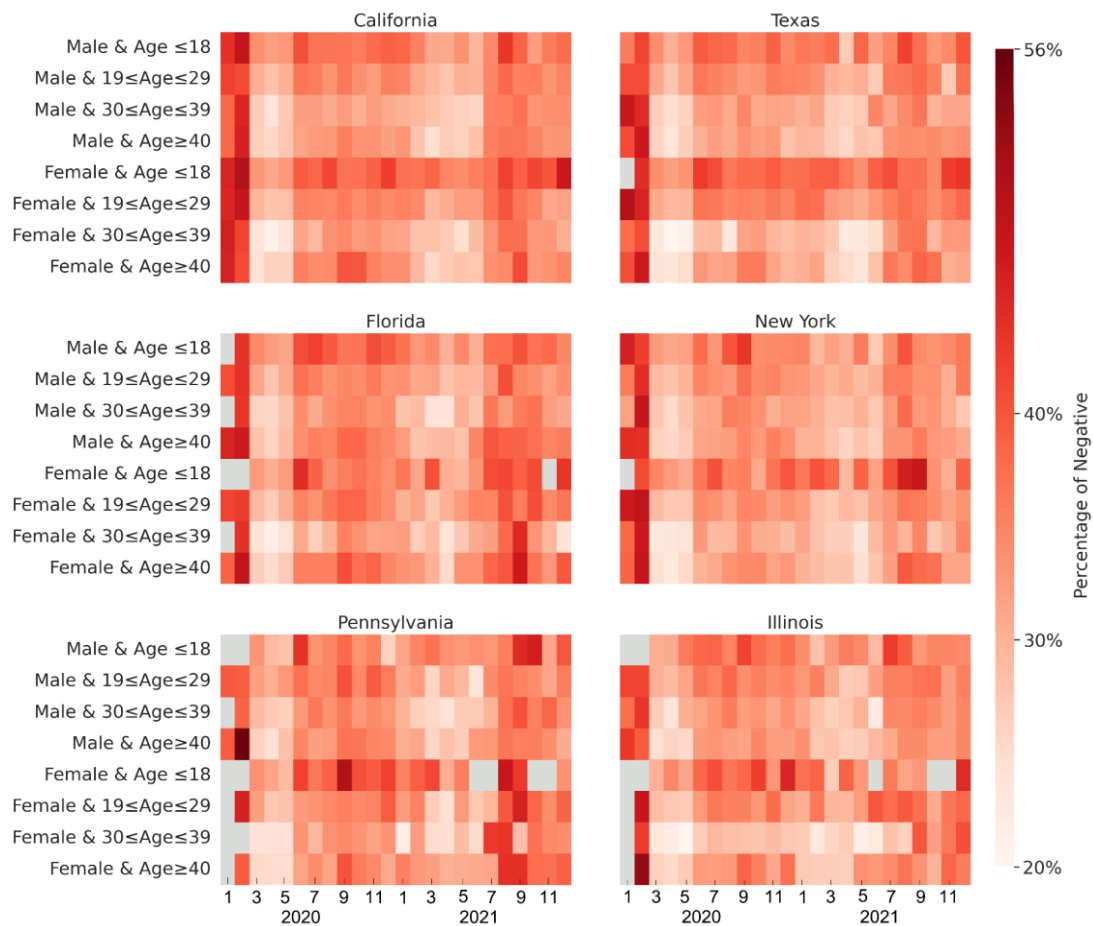

Figure 7. Monthly percentages of negative Twitter users toward COVID-19 in different social groups in the six most populous states.

### 4.4 Sentiment adjusted by demographics (SAD) Index

#### 4.4.1 Temporal trends of SAD Index at the national level

Figure 8 illustrates the temporal trends of the numbers, original rates, and adjusted proportions of users expressing positive, neutral, and negative sentiments toward COVID-19 in 2020 and 2021 in the U.S. The monthly numbers of Twitter users discussing COVID-19 ranged from 9,004 to 285,145, with an average value of 83,819. Prior to March 2020, there was minimal discussion about COVID-19 on Twitter, despite reports of outbreaks occurring outside the U.S. The largest number of Twitter users tweeting about COVID-19 was observed in March 2020, when the World Health Organization (WHO) declared the novel coronavirus outbreak as a pandemic, and the U.S. declared a national emergency. Subsequently, the number of Twitter users concerning about COVID-19 decreased to 99,170 in June 2020 and followed by an overall downward trend with four minor peaks in July 2020, October to November 2020, August 2021, and December 2021.

The subfigure below in Figure 8 depicts the difference between the percentages of Twitter users falling into positive, neutral, and negative sentiment categories before and after adjustment. It was found that the adjusted user percentages were generally similar to the original user percentages across all three sentiment levels. However, there were some differences observed. Adjusted percentages were generally higher than raw percentages for negative sentiment, lower for neutral sentiment, and comparable for positive sentiment.

The adjusted monthly percentages of Twitter user at three sentiment levels reflect the national sentiment towards the pandemic over time. The monthly percentages of negative Twitter users ranged from 26.65% to 44.56% with an average of 33.56%. The monthly percentages of negative Twitter users were high (41.07%) in January and reached their maximum in February 2020. Afterwards, there was a significant decrease in the percentages of negative users, dropping to 26.65% in April 2020, followed by a rebound to 34.68% in June 2020. Throughout the rest of 2020, the percentages remained relatively stable around 35%. In 2021, the percentages of negative Twitter users gradually declined from 33.05% in January 2021 to 27.98% in April 2021 and moderately rose to 37.56% in September 2021. The percentages of negative Twitter users then dropped to 33.48% in November and bounced up to 34.99% in December 2021.

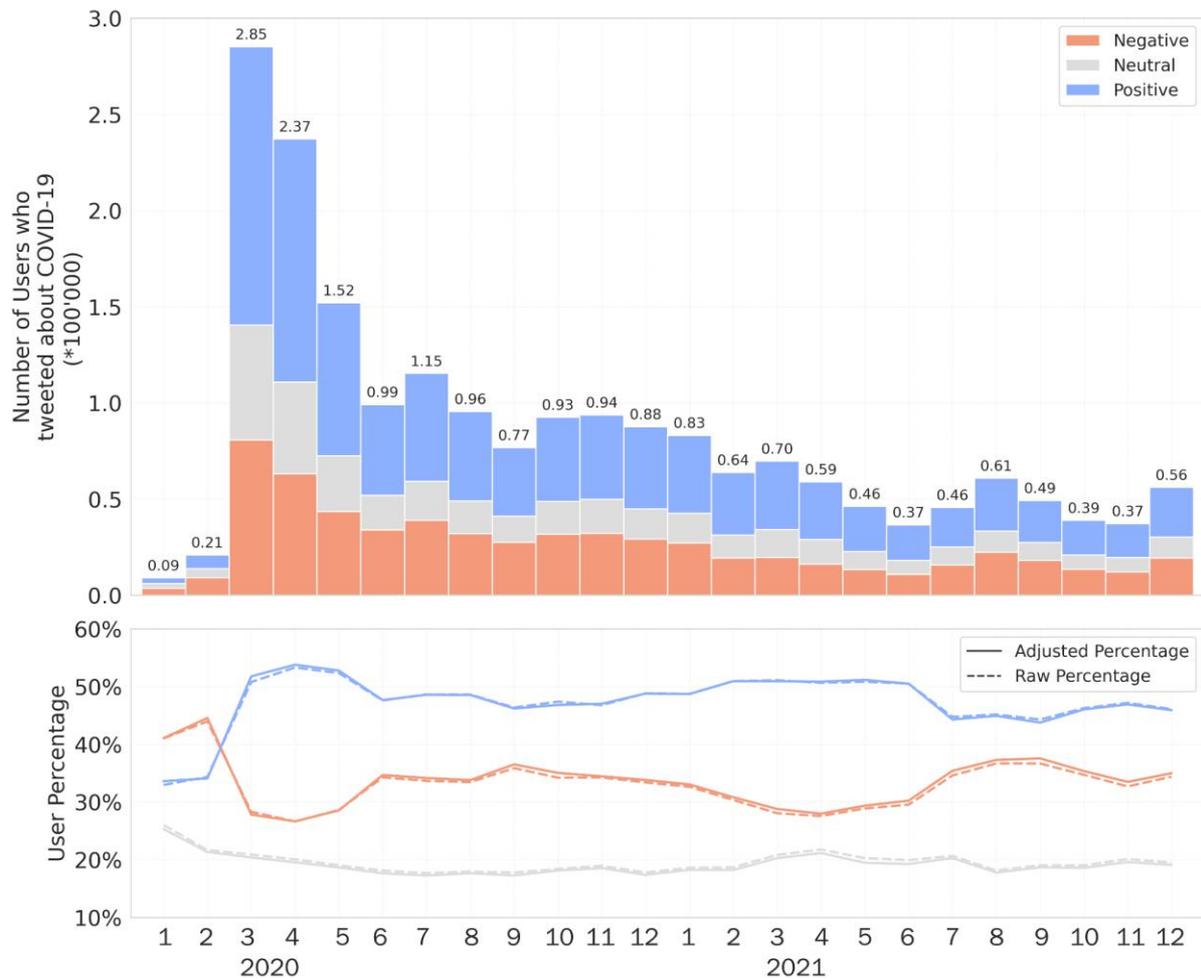

Figure 8. Temporal trends of the numbers, original rates, and adjusted proportions of users expressing positive, neutral, and negative sentiments toward COVID-19 in the U.S. during the period of 2020-2021.

*4.4.2 Spatial disparities of SAD Index at the state level*

Figure 9 displays state-level maps of the unadjusted percentages of Twitter users expressing negative sentiments toward COVID-19 (left) and the SAD index (right). The three lines encompassing each national map represent the percentages of Twitter users categorized into different sentiment levels before and after demographic adjustment. Figure 10 presents a scatter plot showcasing the original percentages of negative Twitter users toward COVID-19 and the SAD index, visualizing the changes in percentages after demographic adjustment at the state level.

The original state-level percentages of negative Twitter users ranged from 20.63% to 32.21%, with a mean value of 28.82%. The state-level SAD index ranged from 22.01% to 33.12%, and the average value was 28.96%. As shown in Figures 9 and 10, the SAD indexes were similar to the original percentages in most states. However, in 29 states and Washington D.C., the SAD indexes exceeded the initial rates of negative Twitter users, with the largest exceeding observed in Vermont (VT), where the value was adjusted from 20.63% to 22.01% (Figure 10). On the other hand, in 21 states, the SAD indexes were lower than the original percentages of negative Twitter users. Texas exhibited the most substantial discrepancy, where the original percentage of 29.21% was adjusted downward to 28.17%.

The adjusted percentages of Twitter users at three sentiment levels revealed the geographical disparities of demographic adjusted sentiment toward pandemic at the state level. The highest SAD index was observed in Wyoming (WY), followed by Montana (MT), Alaska (AK), South Dakota (SD), and New Mexico (NM), with values of 32.98%, 32.35%, 31.85%, and 31.70%, respectively. Vermont showed the lowest SAD index (22.00%), followed by Massachusetts (MA), Nebraska (NE), District of Columbia (DC), and North Carolina (NC), with values of 26.54%, 26.92%, 26.93%, and 27.06%. The adjusted percentages of positive Twitter users ranged from 46.95% to 56.29% with an average value of 52.46%. Vermont had the highest percentage, while Alaska displayed the smallest proportion of users showing negative emotions toward COVID-19. The adjusted percentages of neutral Twitter users ranged from 16.77% to 21.70%, and the mean value was 18.59%.

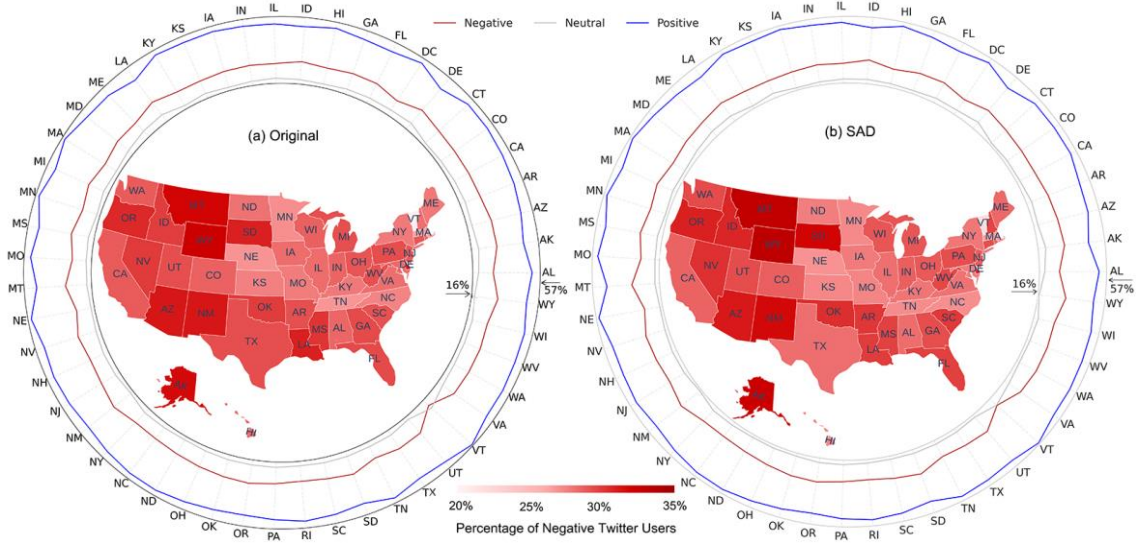

Figure 9. State-level percentages of negative Twitter users toward COVID-19 before and after demographic adjustments.

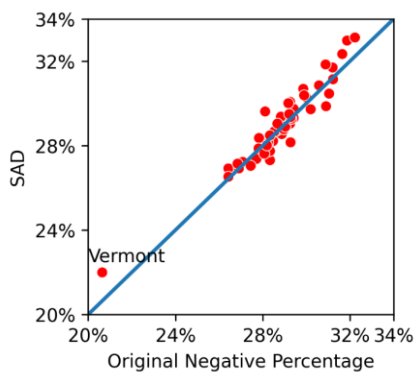

Figure 10. The state-level comparison of original and adjusted percentages of negative Twitter users toward COVID-19.

## 5. Discussion

### *5.1 Significant Implications*

This study has several significant implications. It found that females and teenagers on Twitter displayed higher levels of depressive toward COVID-19. The finding aligns with a scientific brief released by WHO, which indicated that females, particularly those aged between 20 and 24, were more affected by the pandemic than males (Brunier, 2022). The lack of mental health support for the younger generation, school closures, and the uncertain job market for graduates all contribute to increased mental health risks among young people. To address this mental health crisis, targeted support for young people and women is necessary. This can include establishing mental health centers and providing training on work-life balance and effective communication in remote settings.

Second, this study demonstrates the existence of demographic bias in social media data and points out the importance of addressing this bias. Failing to account for such bias can result in overlooking the needs and experiences of specific social groups. The framework developed in this research for demographic adjustment provides a valuable approach for mitigating the bias when conducting analyses using social media data. This framework opens up opportunities for diverse analyses, including assessing public awareness, monitoring human mobility, detecting trending topic, and tracking human behaviors during future pandemics or natural disasters.

Finally, the SAD index proposed in this research provides a demographically adjusted evaluation of public sentiment toward COVID-19. The dataset containing the daily, weekly, and monthly SAD index at the state level is available as a GitHub repository (https://github.com/binbinlinGISer/Sentiment-adjusted-by-demographics-SAD-Index). The SAD index dataset holds great potential for various research applications, such as monitoring mental health in different social groups, analyzing the dynamic causal relationships between humans and pandemics, and informing evidence-based decision-making and policymaking for pandemic control measures.

### *5.2 The associations between public sentiment and COVID-19*

We conducted a correlation analysis between public sentiment and the COVID-19 health impact to illustrate the use scenario of the findings. Figure 11 displays the weekly time series of the SAD index and the case rate in the U.S. during the period of 2020 to 2021. To investigate potential associations between negative public sentiment towards COVID-19 on Twitter and the pandemic's spread, a hypothesis was formulated proposing a positive relationship between the SAD index and case rate.

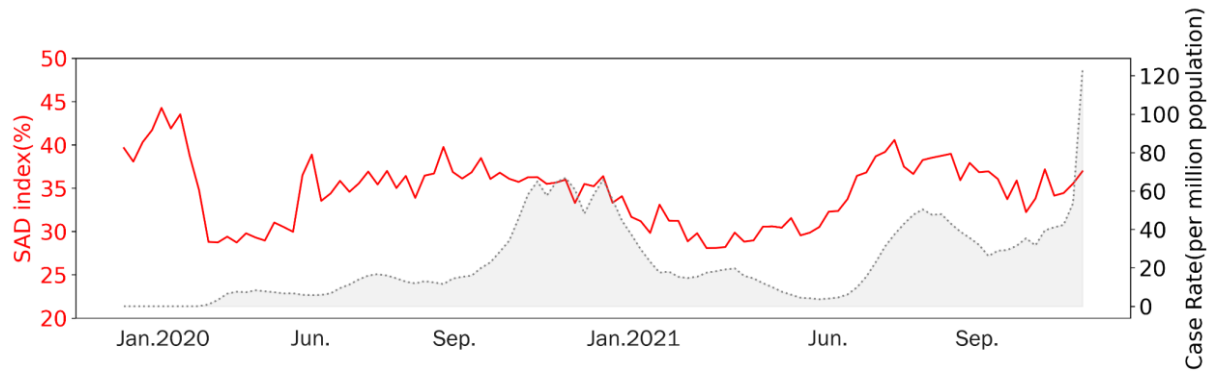

Figure 11. The time series of weekly SAD index and case rate in the U.S. in 2020 and 2021.

The results show that in 2020, there was no statistically significant association between the SAD Index and the case rate, with a Pearson correlation coefficient of -0.03 and a p-value of 0.85. In contrast, for 2021, the results demonstrated a positive and significant association between the SAD Index and the case rate, with a Pearson correlation coefficient of 0.55 and a p-value less than 0.01. These findings suggest that the SAD Index was more strongly influenced by the case rate in 2021 compared to 2020 in the U.S., which could be attributed to various reasons. During the early stages of the pandemic in the U.S., the public sentiment represented by the SAD Index may have been influenced more by news and information about the outbreak in other countries, rather than the actual case rate in the U.S. This could have resulted in a higher level of negative sentiment on social media, despite the relatively low case rate in the U.S. In 2021, the rapid surge in cases caused by the Delta variant led to a sharp increase in cases in a relatively short period, which may have more closely aligned the negative public sentiment towards COVID-19 with the actual case rate in the U.S.

*5.3 Limitations and Future Research*

There are several limitations in this study that necessitate further investigation. The innate shortage of social media data is the non-representations of the entire population. Studies show that vast majority of tweets (97%) are produced by a minority of Twitter users (25%) (McClain et al., 2021). This means that the sentiment of Twitter users cannot represent the emotions of people who do not use Twitter, and the sentiment of Twitter users tweeting about COVID-19 cannot represent remaining silent Twitter users. This challenge is inevitable for social media-based human behavioral analysis.

Moreover, we applied the M3 model in this research to detect the gender and age of Twitter users. Although this model is proved as a more accurate model compared to other models, the accuracies and efficiency of the M3 model for age and gender detection are imperfect, leading to uncertainty and time-consuming processes in detecting the demographic disparities in public sentiment toward COVID-19. Future studies may consider transferring to alternative social media data providing demographic information or constructing a demographic detection model with significant increase in accuracy and efficiency based on state-of-the-art AI technologies to tackle the limitation. Additionally, although the granularities of age and gender categories are sufficient to analyze the sentiment in different social groups, the sentiment disparities of people older than 40 years old, such as middle and old-aged people and elderly people, cannot be distinguished in this study. Nonbinary people were not included in this study, potentially leading us to overlook the sentiments of other gender identities that exist in our society. Therefore, sentiment analysis among more detailed age groups and more diverse gender groups could be taken into consideration in future studies.

Finally, it is challenging to verify the prominence of the SAD index. National Center for Health Statistics (NCHS) and the Census Bureau (https://www.cdc.gov/nchs/covid19/pulse/mental-health.htm) have surveyed residents to collect data and create indicators of anxiety and depression. However, sentiment, mental health, anxiety, and depression are distinct concepts and cannot be compared directly. Mental health is a grand concept, including emotional, psychological, and social well-being (Substance Abuse and Mental Health Services Administration, 2023), while sentiment is an attitude, thought, or judgment prompted by feeling (Wikipedia, 2022). For example, smiling depression is a term for someone living with depression on the inside while appearing perfectly happy or content on the outside (Gunnerson, 2022), such as posting positive tweets. More supporting data from surveys are needed to validate the estimation from the SAD index.

## 6. Conclusion

In this study, the sentiment of COVID-19-related Twitter data in the U.S. was analyzed for the years 2020 and 2021. The study aimed to address two key research gaps. First, it sought to understand the geographical, temporal, and demographic disparities in public sentiment toward COVID-19 as reflected on social media. Second, it aimed to develop strategies for mitigating the demographic bias within social media data to accurately evaluate spatial and temporal patterns of public sentiment toward specific topics or events such as the COVID-19 pandemic. To achieve this, the study utilized an advanced AI-based model to identify demographic biases in Twitter usage and incorporate such bias for estimating public sentiment relevant to COVID-19 using the post-stratification method.

This research has yielded significant findings. First, it reveals the presence of demographic disparities in sentiment toward COVID-19 on Twitter in the U.S., which vary across different spatial and temporal scales. Nationally, female Twitter users and those under the age of 18 consistently exhibited the largest percentages of users showing negative emotions when discussing about COVID-19, indicating their vulnerability to the pandemic. At the state level, the percentages of users showing negative emotions toward COVID-19 varied over time and across states. For instance, in September 2021, female Twitter users aged 30 and above in Florida showed the highest percentage of negative users. Second, by accounting for demographic differences between Twitter users and the general population, the proposed SAD index offers a demographically adjusted evaluation of public sentiment toward COVID-19. The SAD index addresses the underestimation of negativity in 29 states and D.C., particularly in Vermont, and overestimation in 21 states. Third, according to the SAD index, Twitter users in different states exhibited varying sentiments toward the pandemic. Wyoming had the highest proportion of negative sentiment towards COVID-19, while Vermont had the highest proportion of users expressing overall positive emotions. Lastly, the increase in COVID-19 cases appears to be a significant factor triggering the rise in negative sentiment towards COVID-19 in the U.S. in 2021.

The adjustment framework of demographic bias of social media data and the proposed SAD index offer valuable insights into diverse applications. This study reveals the overlooked voices of digitally marginalized populations, shedding light on the emotion changes and possible needs of these populations during the pandemic. The framework can be applied to adjust the demographic biases when monitoring other human behaviors, such as public awareness and human mobility. Researchers can leverage the SAD index to investigate the interplay among changes of public sentiment, human mobility, lockdown policies, and COVID-19 health impacts. Finally, the findings can identify vulnerable social groups during health crises and inform policymakers as well as individuals to better navigate future pandemics.

**Acknowledgements**

This study is based on work supported by the Data Resource Develop Program Award from the Texas A&M Institute of Data Science (TAMIDS). Any opinions, findings, and conclusions or recommendations expressed in this material are those of the authors and do not necessarily reflect the views of the funding agencies.

**Data Availability Statement**

The data used in this research were derived from the following resources available in the public domain: Twitter Application Programming Interface (API) for Academic Research (https://developer.twitter.com/en/products/twitter-api/academic-research), U.S. Census Bureau (https://www.census.gov/programs-surveys/popest/technical-documentation/research/evaluation-estimates/2020-evaluation-estimates/2010s-national-detail.html), M3 model repository (https://github.com/euagendas/m3inference), and COVID-19 Data Repository by the Center for Systems Science and Engineering (CSSE) at Johns Hopkins University (https://github.com/CSSEGISandData/COVID-19). The SAD index dataset generated in this study is available as a GitHub repository (https://github.com/binbinlinGISer/Sentiment-adjusted-by-demographics-SAD-Index).

**Disclosure statement**

No potential conflict of interest was reported by the authors.

**Figures**

Figure 1. Framework of Twitter data mining for demographic-adjusted public sentiment evaluation.

Figure 2. Comparison of demographic structures between Twitter users discussing COVID-19 and the actual population in the U.S.

Figure 3. Percentages of negative, neutral, and positive Twitter users in eight social groups in the U.S.

Figure 4. Temporal trends of percentages of negative Twitter users in different ages and gender groups in the U.S.

Figure 5. Numbers and ratios of Twitter users who tweeted about COVID-19 by gender and age at the state level in the U.S. during the period of 2020 to 2021.

Figure 6. Percentages of negative Twitter users toward COVID-19 within various social groups at the state level across the U.S. during the period of 2020 to 2021.

Figure 7. Monthly percentages of negative Twitter users toward COVID-19 in different social groups in the six most populous states.

Figure 8. Temporal trends of the numbers, original rates, and adjusted proportions of users expressing positive, neutral, and negative sentiments toward COVID-19 in the U.S. during the period of 2020-2021.

Figure 9. State-level percentages of negative Twitter users toward COVID-19 before and after demographic adjustments.

Figure 10. The state-level comparison of original and adjusted percentages of negative Twitter users toward COVID-19.

Figure 11. The time series of weekly SAD index and case rate in the U.S. in 2020 and 2021.